# Design of Digital Response in Enzyme-Based Bioanalytical Systems for Information Processing Applications

**Sergii Domanskyi**[a] and **Vladimir Privman**[a,*]

[a]Department of Physics, Clarkson University, Potsdam, New York 13699, USA

[*]Corresponding author: e-mail privman@clarkson.edu, telephone +1-315-268-3891

## ABSTRACT

We investigate performance and optimization of the "digital" bioanalytical response. Specifically, we consider the recently introduced approach of a partial input conversion into inactive compounds, resulting in the "branch point effect" similar to that encountered in biological systems. This corresponds to an "intensity filter," which can yield a binary-type sigmoid-response output signal of interest in information and signal processing and in biosensing applications. We define measures for optimizing the response for information processing applications based on the kinetic modeling of the enzymatic reactions involved, and apply the developed approach to the recently published data for glucose detection.

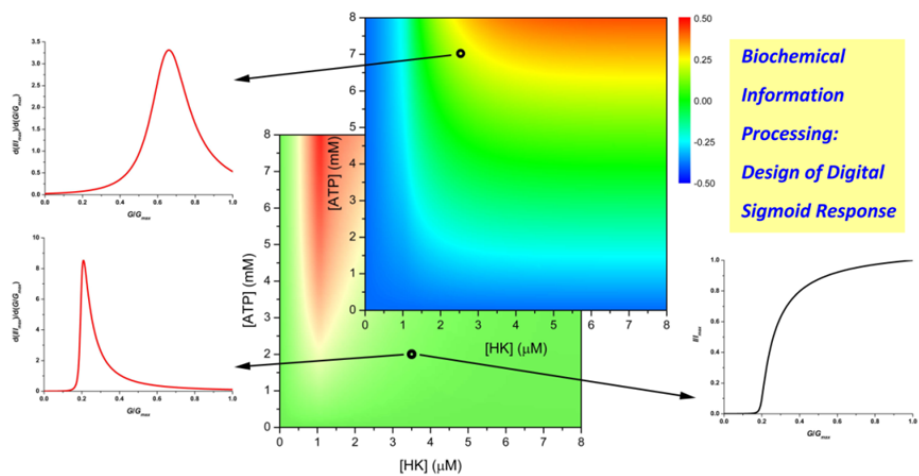

## INTRODUCTION

Information processing based on biomolecular processes has recently received considerable attention as a promising area of research in "digital" sensor and actuator design, logic systems, and novel approaches to interfacing and computing in biotechnology.[1,2] Specifically, enzyme-catalyzed reactions have been used in new biomolecular logic systems being developed for novel diagnostic applications.[1-6] Biomolecular logic has been studied for future biomedical and diagnostic applications requiring analysis of biomarkers characteristic of different pathophysiological conditions indicative of illnesses or injury.[7-10] Such biomolecular logic gates and networks are "binary" (digital), giving two distinct outputs, as detailed shortly. They can provide new functionalities for analytical purposes, offering a new class of biosensors which can process several input signals and generate a binary output of the alert type: YES/NO, i.e., biosensor-bioactuator "Sense/Act/Treat" systems.[11-14] This approach has already been successfully applied to analyze sets of biomarkers indicative of various kinds of trauma.[15,16] By binary (digital) in these applications we mean processing which identifies well-defined values or ranges of values corresponding to the reference **1** or **0** (YES/NO, Act/Don't Act) levels of the signals.[17] Standard logic gates, such as AND, OR, XOR, INHIB, etc.,[5,18-22] and several-gate biomolecular networks[23-27] have been realized, which process information mimicking digital electronics designs.

Recently, efforts have been devoted to understanding the control of noise in functioning of biomolecular gates used as network elements.[5,17,24] A particularly promising approach has been to modify the typical biomolecular reaction response (output as a function of the inputs) from convex to sigmoid, the latter common in natural systems.[5,28,29] Biomolecular filters based on different mechanisms have been devised, ranging from the use of enzymes with substrates that have self-promoter properties,[30] to redox transformations,[31] pH control by buffering,[32] and utilization of competing enzymatic processes.[33] These developments have built on earlier approaches to understand or realize a sigmoid or all-or-none (ON/OFF, YES/NO) response in natural or synthetic biological and biochemical systems.[34-37]

The latest efforts in biochemical filter designs have been focused on "intensity filtering" whereby a fraction of the output[5,21,31] signal or that of an input signal is neutralized[33] by an added chemical process or converted into one of the intermediate reagents in the cascade. These concepts will be further defined and explained in the later sections. The approaches are interrelated especially when the realized gates are considered for networking (so that outputs then become inputs to other gates). The partial output neutralizing approach has been successfully applied to systems of interest in applications[10] as well as yielded realizations[38,39] of double-sigmoid (means, with "filtering" properties with respect to both inputs) AND and OR logic gates.

Intensity-filtering based on partial input neutralization has not been theoretically analyzed for optimizing the resulting output signal for information processing. The present work develops such a modeling approach. In the next section we describe the studied system for which experimental data were obtained in Ref. 33. We introduce a kinetic description of the enzymatic processes involved which is simplified enough to limit the number of fitted rate constants. At the same time, it allows us to identify those parameters of biocatalytic processes which could be adjusted to control the quality of the realized sigmoid response. The following section presents the results of the actual data analysis and introduces and evaluates quality measures of the sigmoid response obtained. Specifically, we consider the steepness and symmetry of the sigmoid curve, as well as the issue of avoiding too much signal loss due to the added "intensity filtering." We propose a possible range of the system parameters to get an optimal response. The final section offers concluding discussion and summary.

**SIGMOID RESPONSE AND ITS MODELING IN TERMS OF ENZYMATIC KINETICS**

We consider enzymatic processes for signal processing applications. For example, the system realized in Ref. 33, corresponds to signal transduction: the simplest "identity" single-input to single-output "logic gate." For "digital" signal/information processing, we usually focus on reference **0** and **1** values. Generally, these "logic values" (or ranges) of inputs are determined by the application, and logic **0** needs not necessarily be at the physical zero. In the present case[33] the input is glucose in solution, the initial concentration of which can be varied. We take the model values 0 mM and 10 mM, see Ref. 33, for the **0** and **1** inputs, respectively.

Here "signal processing" was biocatalyzed by an electrode-attached enzyme glucose oxidase, via the reaction which also consumes the dissolved oxygen, resulting in the oxidation of glucose yielding gluconic acid. The output was measured[33] at the "gate time" $t_g = 180$ sec as the current which is due to the two-elementary-units charge transfer at the electrode per reaction cycle. In Figure 1 (the figures are appended at the end of this preprint staring on Page 15), the current, $I(t_g)$, normalized per its maximum value $I_{max}(t_g)$ for the largest glucose input, $G_{max}$ = 10 mM, in plotted vs. the glucose input. The data are taken from Ref. 33, whereas our model fit is detailed later.

The binary values **0** and **1** are used to define the YES/NO signals. However, for network functionality/scalability, specifically for evaluating the effects of noise[5,40] in the signals, we have to consider the shape of the whole response curve, e.g., Figure 1, i.e., the output current vs. the input glucose concentration near the logic points and also generally over the whole interval between the binary **0** and **1** reference values and somewhat beyond. Typical response curves of biocatalytic reactions are convex, as in Figure 1. As described in the Introduction, recently there has been a lot of interest in converting these to sigmoid, which offers advantages in noise handling, because small or zero slope at the logic points results in suppression of noise in the input as it is transferred to the output by the "logic gate" function.

Here we consider the approach realized in Ref. 33, of redirecting (consuming) a fraction of the input (glucose) in a competing chemical process, but one that only can use a limited amount of glucose. This modified system can yield sigmoid behavior.[33] Enzyme hexokinase was added to the working solution, and adenosine triphosphate (ATP) was introduced in limited amounts as compared to the maximum 10 mM of glucose, to "switch on" the filtering effect. Indeed, the process biocatalyzed by hexokinase consumes glucose without contributing to the output current. This makes the response of the main process, which generates the current at the electrode, sigmoid. The corresponding experimental points from Ref. 33 and model fit are shown in Figure 2.

Instead of consuming[33] a fraction of the input to achieve sigmoid response, a mechanism of consumption or redirection of the output product is sometimes used.[5,10,32] The two approaches are related because when processes are networked, outputs of some steps become inputs for others. In all cases of such "intensity filtering," there is the tradeoff that output signal intensity can be reduced. Thus, careful optimization is required, involving the analysis of possible sources of noise and its handling.[40,41] Criteria for such optimization for the present type of systems will be addressed in the next section.

We aim at developing models which go beyond an entirely phenomenological data fitting with properly shaped (convex or sigmoid) curves, because we want to explicitly identify and model the dependence on those parameters which could be controlled to optimize the system's response. Phenomenological approaches[42] include the Hill function fitting.[33] Here we use a different approach based on rate-equation modeling of the key steps of the enzymatic processes which allows us to identify the concentrations of hexokinase and ATP as parameters to change to significantly improve the sigmoid response.

In our modeling we focus on few key processes for each enzymatic reaction. The main reason has been that we want few adjustable parameters, as suitable for the typical low-quality data sets available in this field. If we introduce more parameters then the conformity with the experimental data might be better, but the extrapolative reliability of the model will be lost. Thus, only enough adjustable parameters are used to allow a schematic overall trend description of the response curves such as those shown in Figures 1 and 2.

Let us first consider the enzymatic process for glucose oxidase (GOx) only, without "filtering." We focus on the following process steps and their rates:

$$E + G \xrightarrow{k_1} C \xrightarrow{k_2} E + \cdots . \tag{1}$$

Here $E$ denotes the concentration of GOx, and $G$ that of glucose. The intermediate complex, gluconolactone, is produced in the first step, with concentration $C$. For glucose, unlike some other possible substrates for GOx, the first step of the reaction can usually be assumed practically irreversible.[43-46] It is important to emphasize that we do not aim at a detailed kinetic study of the enzymatic reactions involved. As pointed out earlier, we seek a simple, few-parameter description of the response curve of the experiment reported in Ref. 33. We ignore the kinetics of all the other reactants, input or product, except for the rate equation for $C(t)$,

$$\frac{dC(t)}{dt} = k_1 G(t)E(t) - k_2 C(t) , \qquad (2)$$

which should be solved with $E(t) = E(0) - C(t)$. Indeed, we need this quantity only, because the current is proportional to the rate of the second reaction in Eq. (1),

$$I(t) \propto k_2 C(t) , \qquad (3)$$

i.e., our output is $I(t_g) \propto C(t_g)$.

Without the hexokinase (HK) part of the process, we can assume that the GOx reaction at the electrode practically does not consume glucose: $G(t) = G(0)$. This assumption is appropriate for electrochemical designs for glucose sensing.[47-49] For our modeling here, we also assume that in all the cases the oxygen concentration is constant (and therefore is absorbed in a rate constant), ignoring the fact that for the largest glucose concentrations some corrections might possibly be needed due to oxygen depletion at the electrode.[33] With these assumptions, Eq. (2) can be solved in closed form,

$$C(t) = \frac{k_1 E(0) G}{k_1 G + k_2} \left[1 - e^{-(k_1 G + k_2)t}\right] . \qquad (4)$$

Here the initial (and later remaining constant) value of $G$ is the input, varying from 0 to $G_{max}$. For logic-gate functioning analysis of such processes, one frequently uses the logic-range variables which vary in [0,1], here

$$x = \frac{G(0)}{G_{max}}, \quad y = \frac{I(t_g;G(0))}{I_{max}} = \frac{C(t_g;G(0))}{C(t_g;G_{max})}, \tag{5}$$

where $I_{max} = I(t_g; G_{max})$. The slope of $y(x)$ near the logic point values 0 and 1, determines the noise transmission factors.[5,17]

The data in Ref. 33 were given as the values of $y$ for several inputs, $G(0)$. Without the filter process, least-squares fit of these data in our case yielded the estimates $k_1 \cong 80$ mM$^{-1}$s$^{-1}$, $k_2 \cong 60$ s$^{-1}$. We point out, however, that these estimates are rather imprecise, as indicated by the numerical fitting procedures. Indeed, here the rate constants are large in the sense that the dimensionless combinations $k_2 t_g$ and $k_1 G_{max} t_g$ are both much larger than 1. This is consistent with other estimates of these rate constants for GOx with glucose as a substrate.[44-46,50] The dependence of the scaled variable $y$ on $G = G(0)$,

$$y = \frac{\frac{k_1 E(0) G}{k_1 G + k_2}\left[1 - e^{-(k_1 G + k_2)t_g}\right]}{\frac{k_1 E(0) G_{max}}{k_1 G_{max} + k_2}\left[1 - e^{-(k_1 G_{max} + k_2)t_g}\right]} \approx \frac{G\left(G_{max} + \frac{k_2}{k_1}\right)}{G_{max}\left(G + \frac{k_2}{k_1}\right)}, \tag{6}$$

is then to a good approximation only controlled by the ratio $k_2/k_1$, for which a relatively precise estimate is possible, $k_2/k_1 = 0.75 \pm 0.02$. The quality of the fits such as that shown in Figure 1, is not impressive, but this is largely due to the noise in the data and here is similar to the situation with the more phenomenological Hill-function fitting reported in Ref. 33.

With the filter process added, in the presence of HK, of concentration denoted $H(t)$, and ATP (adenosine triphosphate), of concentration $A(t)$, the concentration of glucose will not remain constant but will be depleted. Data are then available[33] for several initial values of $A(0)$. The latter are all smaller than $G_{max}$. Therefore, in order to limit the number of adjustable parameters we will only consider that pathway of the HK biocatalytic process[51,52] in which glucose is taken in as the first substrate, to form a complex of concentration $D(t)$. We again take a very simplified scheme for the HK activity, ignoring a possible reversibility of the complex formation and other process details,[54-51]

$$H + G \xrightarrow{k_3} D + \cdots, \quad D + A \xrightarrow{k_4} H + \cdots. \qquad (7)$$

This simplified approach yields only two adjustable parameters which enter the rate equations that determine the time-dependence of glucose to use in Eq. (2) for calculating *C*(*t*),

$$\begin{aligned}
\frac{dG}{dt} &= -k_3 HG, \\
\frac{dH}{dt} &= -k_3 HG + k_4 DA, \\
\frac{dD}{dt} &= k_3 HG - k_4 DA, \\
\frac{dA}{dt} &= -k_4 DA.
\end{aligned} \qquad (8)$$

Note that the two middle equations can be made into one by using $D(t) + H(t) = H(0)$. The available data were for $H(0) = 2$ µM. The resulting system of differential equations was programmed for a numerical solution, and the data available for four initial nonzero ATP concentrations were fitted to yield the estimates $k_3 = 14.3 \pm 0.7 \text{mM}^{-1}\text{s}^{-1}$, $k_4 = 8.1 \pm 0.4 \text{ mM}^{-1}\text{s}^{-1}$. The earlier estimate for $k_2/k_1$ was used to obtain these values.

**SIGMOID RESPONSE OPTIMIZATION**

In order to enable fault-tolerant[22-24] information processing with considered gates, when they are connected in a network,[55,56] parameters must be chosen to make the analog noise amplification minimal or avoid it (which is accomplished by filtering). There are several sources of noise in the functioning of biochemical reaction processes when they are used as "gates" for signal and information processing. In addition to the noise resulting from the imprecise and/or noisy realization of the expected response curve, *y*(*x*), the latter evident in the data shown in Figures 1 and 2, there is also noise transmission from input to output. In biochemical environments the noise in the inputs is typically quite large.[1,2,5,17,57-60] As mentioned, avoiding this "analog noise" amplification during signal processing is paramount to network stability. For larger networks, additional consideration of "digital" errors[17] is required, but here we focus on the single-processing-step design.

Unless the input noise levels are very large or the response curve, here $y(x)$, has some non-smooth features near the logic point $x$ values 0 or 1, then the noise transmission factor is simply given by the slope of the curve $y(x)$ near each of the two logic points. Filtering makes both much smaller than 1, cf. Figures 1 and 2. For best results, the filtering response-curve shape should be symmetrically positioned and steep. This, however, should not be done at the expense of losing much of the intensity of the output signal in terms of its actual range of values, here equal $I_{max}$, as opposed to the scaled variable $y$ which always varies in [0,1]. Indeed, loss of intensity amplifies the *relative level* of noise from all the sources discussed above.

The input values are set by the gate usage and typically cannot be adjusted. However, we can select other parameters to control the filtering quality, as already initiated in Ref. 33. Here we formulate quantitative criteria for such a gate-quality evaluation. Note that within the assumed regime of functioning, the shape of $y(x)$ does not depend on $E(0)$. However, other "gate machinery" (means, not input or outputs) initial chemical concentrations can be varied. We consider the adjustment of both $H(0)$ and $A(0)$. Other modifications can of course be contemplated, such as changing the physical or chemical conditions (which affects the rates of various processes) or limiting the supply of oxygen.[61]

A sigmoid curve has a peaked derivative. In order to achieve the response curve as symmetric as possible we consider the position of the peak of $y'(x)$. In enzymatic processes, sigmoid response-curves is typically not symmetrical with respect to the inflection region. We define the width of the peak of the derivative, by the difference of the two values $x_2 - x_1$ defined by $y'(x_{1,2}) = 1$. The middle-point position of the peak will be taken at $(x_2 + x_1)/2$. Figure 3 shows three different illustrative sigmoid response curves, as well as their derivatives calculated within the model and with the parameter values discussed in the preceding section. Figure 4 presents a contour plot of the deviation of the middle-point peak position from 1/2. Our aim is to get it rather close to 1/2 without compromising the other gate-quality criteria. One of these is analyzed in Figure 5, which maps out the width of the peak, which we would like to be as small as possible.

Yet another criterion is that of avoiding the loss of the actual signal intensity, to the extent possible. Since enzymatic processes usually approach saturation at large inputs, here this can be defined as the fractional loss:

$$1 - \frac{I_{max}(H(0)>0, A(0)>0)}{I_{max}(H(0)=0, A(0)=0)} = 1 - \frac{I(t_g, G_{max})_{H(0)>0, A(0)>0}}{I(t_g, G_{max})_{H(0)=0, A(0)=0}}. \tag{9}$$

This quantity is shown in Figure 6 as the percentage value. Figures 4, 5, 6 span values safely within the experimentally realizable ranges of the considered control quantities, $H(0)$ and $A(0)$. Consideration of Figures 4 and 5 suggests that the peak can be made optimally centrally positioned and narrow, by selecting approximately $H(0) = 4\,\mu M$ and $A(0) = 4\,mM$. The optimal choices correspond to the regions marked by the white ovals in the figures. The loss of intensity is usually present for this type of filtering. However, it can be tolerated if percentage-wise it is comparable to (or smaller than) the degree of noise otherwise present in the output. The approximately 5% loss level in the oval-delineated region (see Figure 6) is therefore acceptable. Our optimal sigmoid response shape and its derivative are shown as curves (a) in the panels of Figure 3. While not symmetrical, the response curve is centrally positioned and rather narrow. The derivative of the output signal in regions $0 \leq x \lesssim 0.37$ and $0.63 \lesssim x \leq 1$ is less than 1, see Fig. 3: bottom panel, curve (a). Thus, in these two input regions, each extending $\sim 37\%$ from the logic points 0 and 1, the errors in the input will not be amplified.

**CONCLUSIONS**

The present work develops a modeling approach to biochemical "intensity filtering." We explored in details the recently realized experimental system and results, Ref. 33. Specifically, we established criteria for optimizing the bioanalytical response, and illustrated their use for the particular experiment considered. Different physical or chemical conditions can be changed to impact enzymatic processes, and we demonstrated how our system's response changed when the initial concentration of two "filter process" chemicals are varied: hexokinase and ATP. The developed criteria are quite general and can be applied to other systems contemplated for information and signal processing, and for biosensing, with biomolecular processes.

## ACKNOWLEDGMENT

We wish to thank Profs. V. Gorshkov and E. Katz, and Drs. J. Halámek and O. Zavalov for helpful input and discussions, and acknowledge funding by the National Science Foundation (grants CBET-1066397 and CCF-1015983).

**FIGURES AND CAPTIONS**

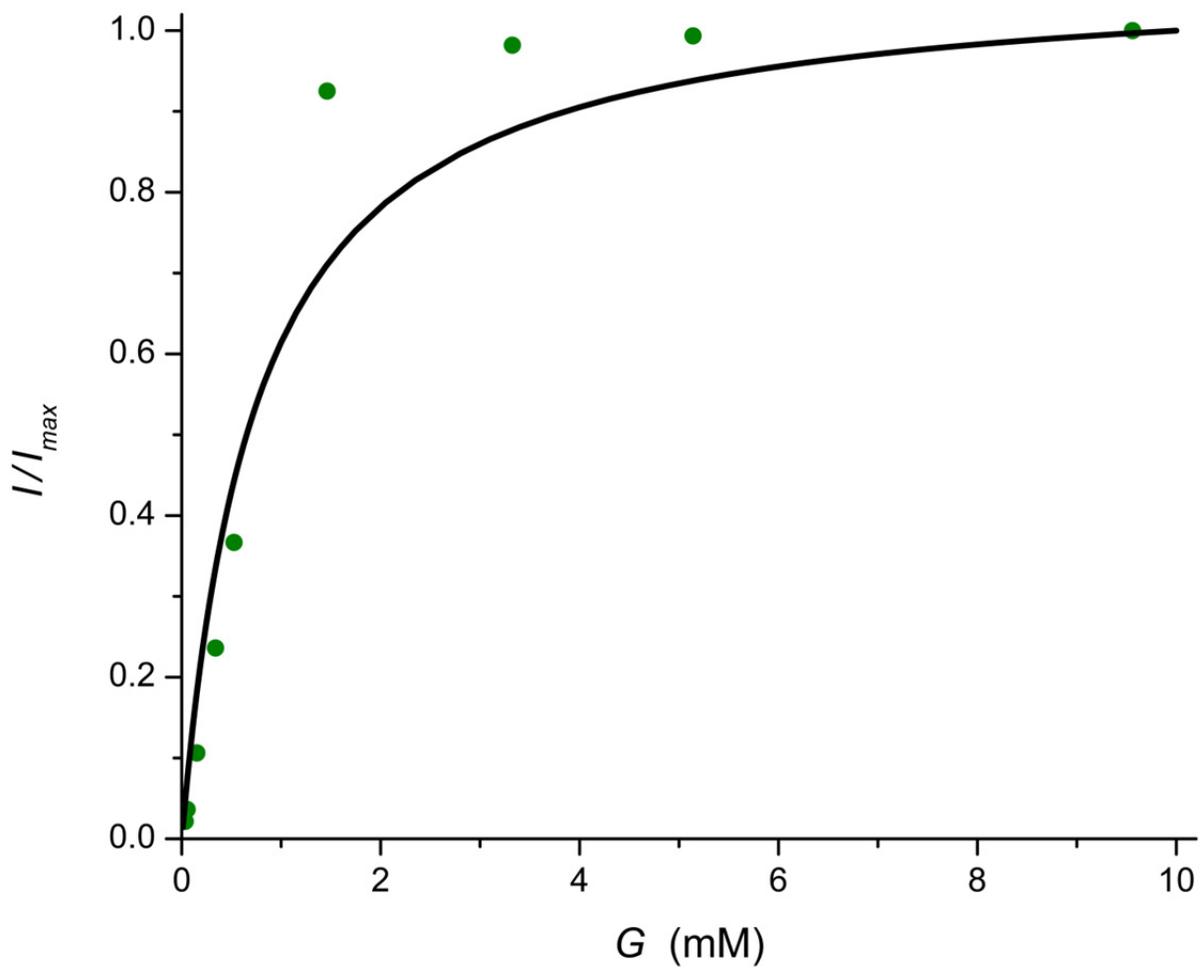

**Figure 1.** Experimental data[33] (circles) for the normalized current and our numerical model (line) without the "filter" process, with the fitted parameters as described in the text.

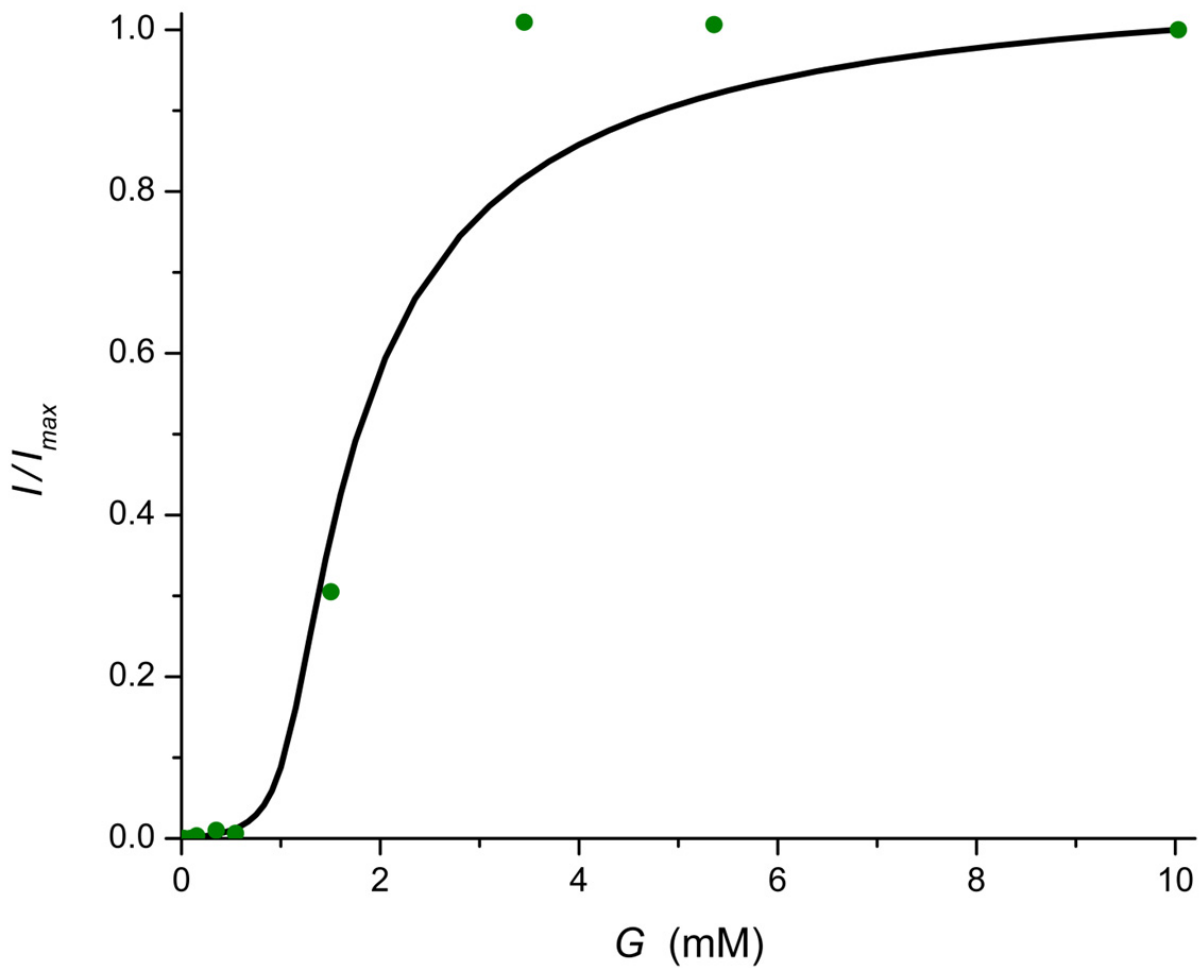

**Figure 2.** Experimental data[33] (circles) for the normalized current and our numerical model (line), with the fitted parameters given in the text. The process here is the same as in Figure 1, but with added hexokinase (2 μM) and the initial concentration of ATP, 1.25 mM, a fraction of the maximum initial glucose concentration, 10 mM.

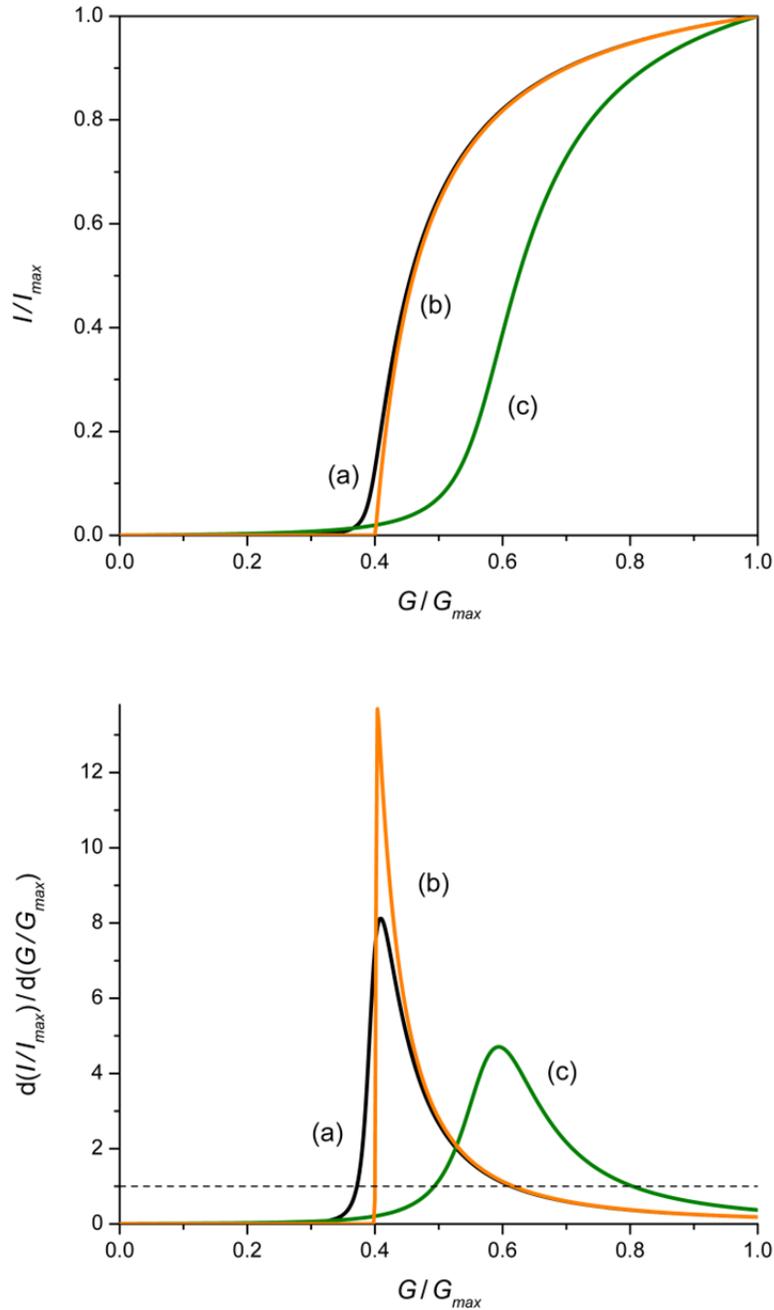

**Figure 3.** Examples of sigmoid curves (top panel) and their derivatives (bottom panel) for three different selections of the parameters used to control the response: (a) $HK = 4\,\mu M$ and $ATP = 4\,mM$; (b) $HK = 8\,\mu M$ and $ATP = 4\,mM$; (c) $HK = 3\,\mu M$ and $ATP = 6\,mM$. The values (a) correspond to the center of the optimal range as described in the text. The dashed line indicates the level at which the width of the peak of the derivative is measured.

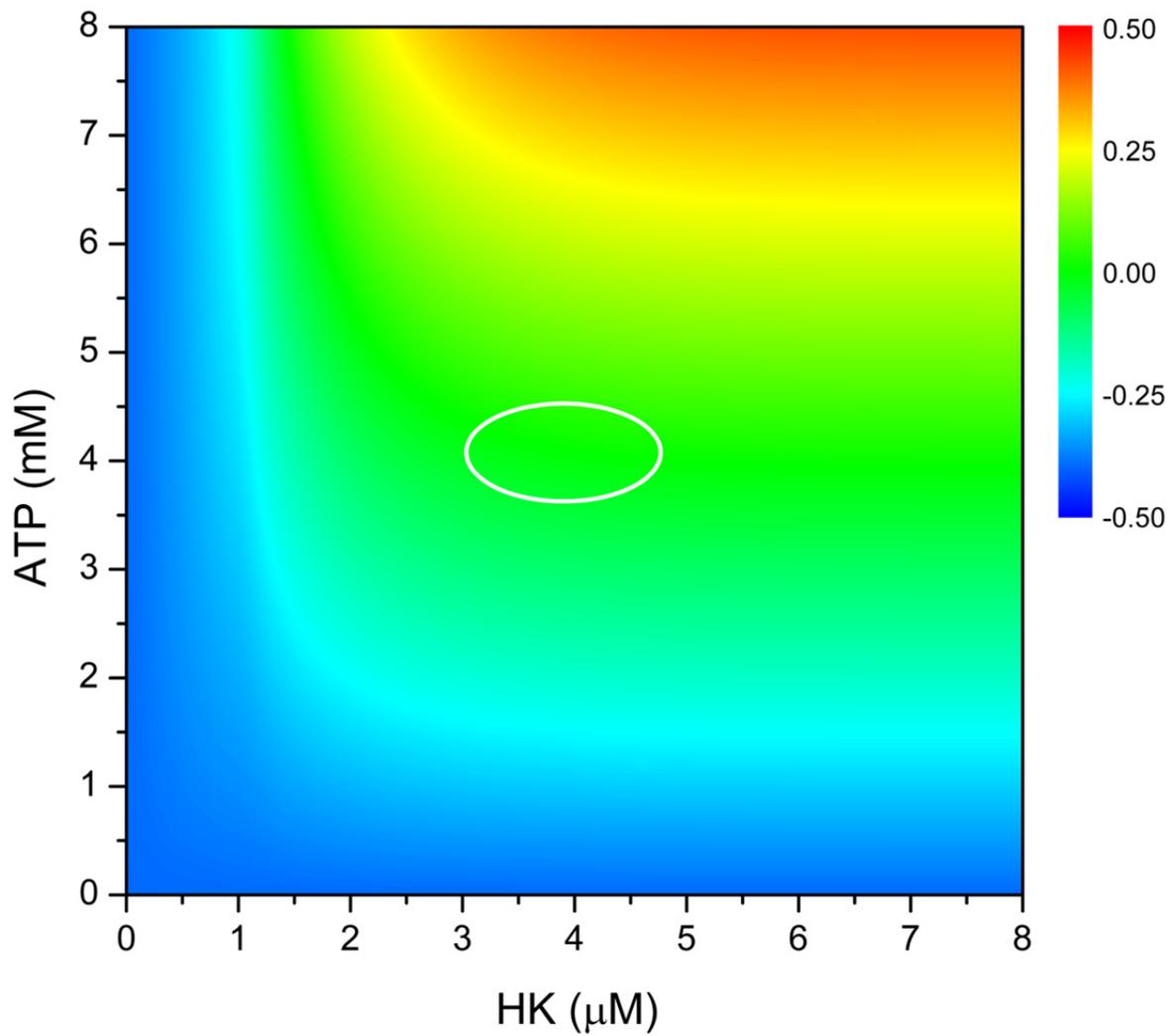

**Figure 4.** Contour plot of the deviation of the middle-point of the peak location from 1/2, i.e., $(x_2 + x_1 - 1)/2$. The optimal values are as small as possible (green color). The oval defines the best choice of the parameters considering the other criteria for optimizing the response: see text.

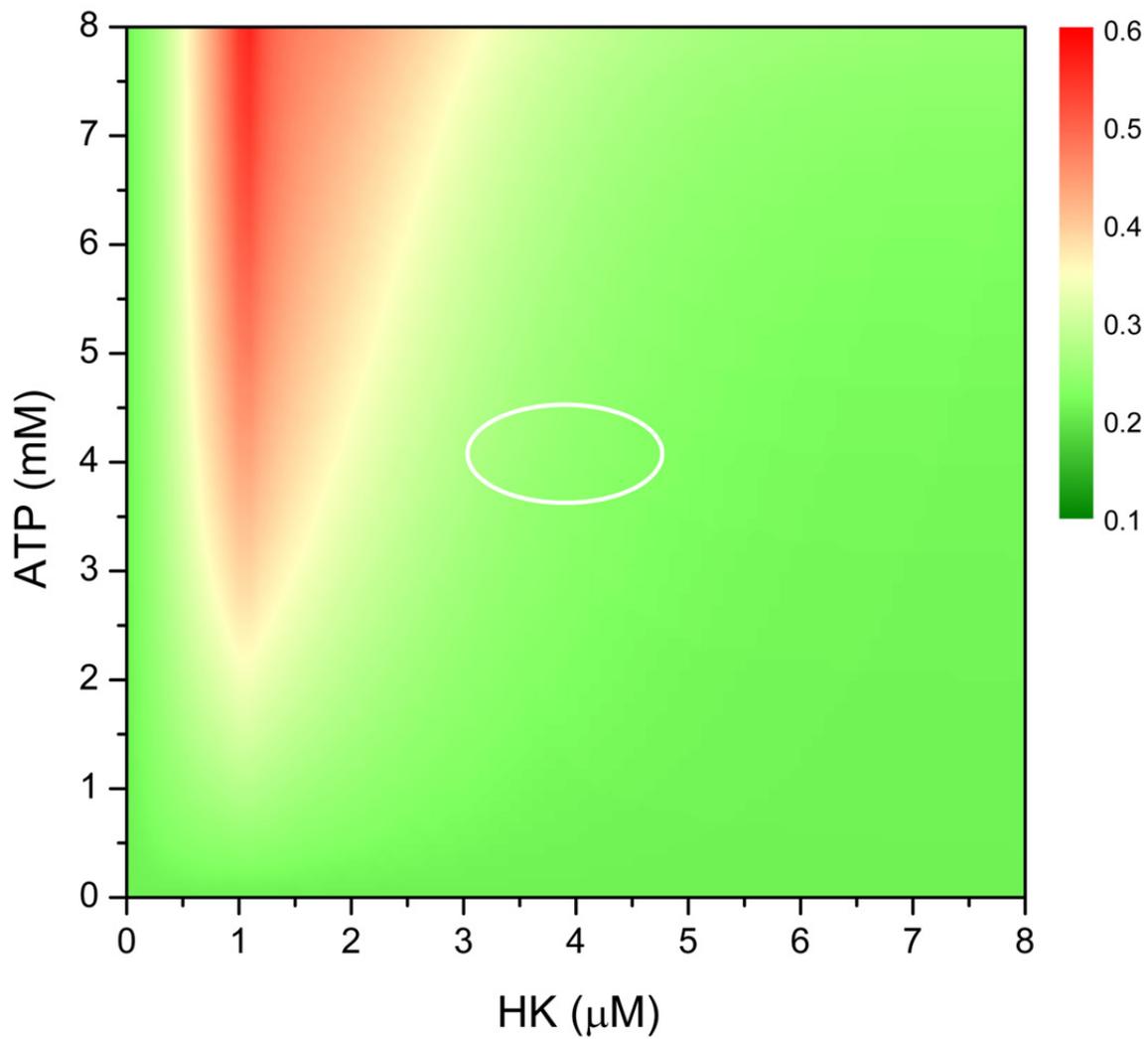

**Figure 5.** Contour plot of the width of the peak. The optimal values are as small as possible (the green shades). The oval defines the best choice of the parameters considering the other criteria for optimizing the response: see text.

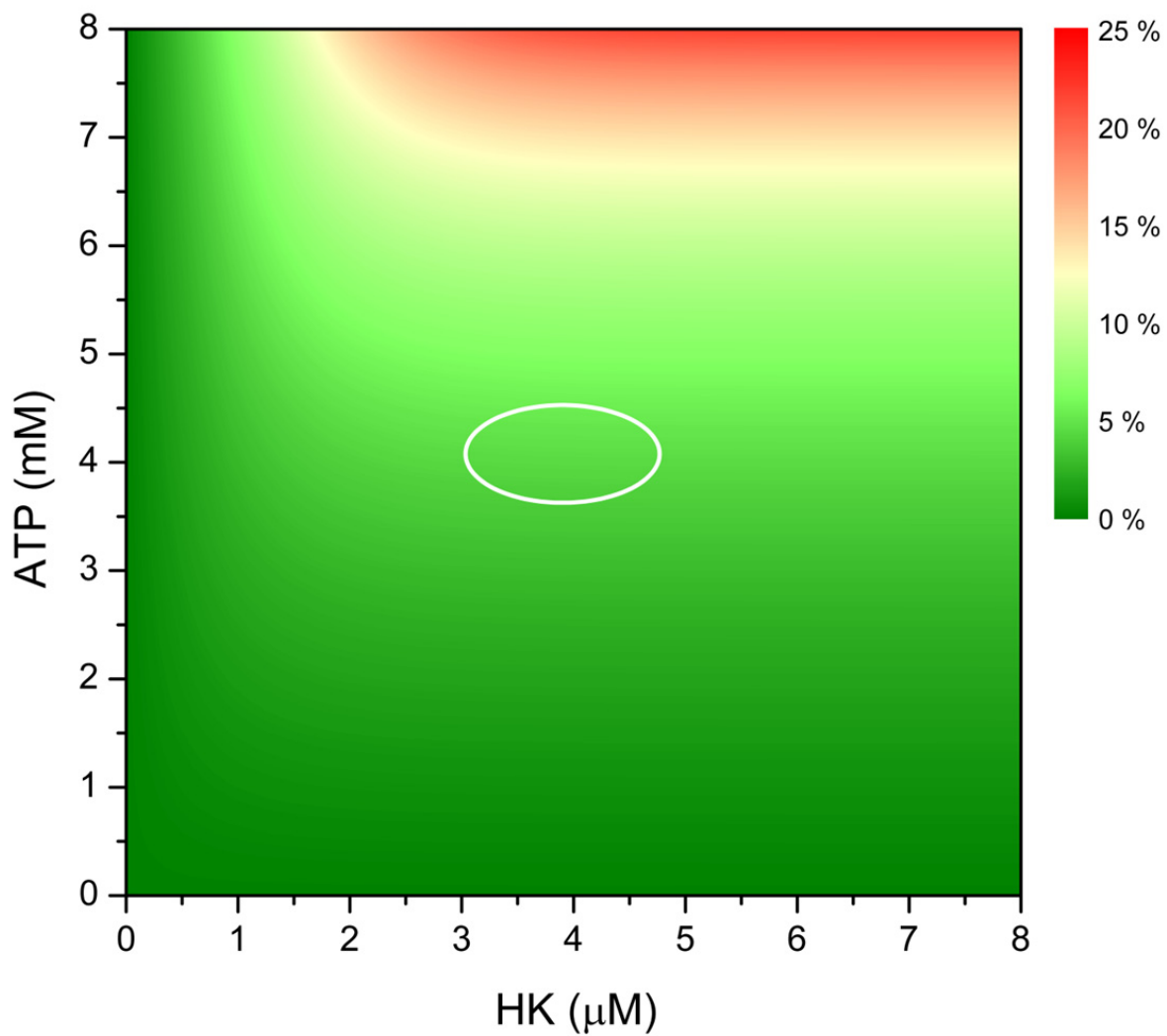

**Figure 6.** Contour plot of the measure of the loss of the output signal intensity, Eq. (9). This measure should be minimized (green color) without compromising the other gate-quality criteria. The oval defines the best choice of the parameters considering the other two criteria: see text.